\documentclass[letterpaper, 10 pt, conference]{ieeeconf}  

\IEEEoverridecommandlockouts                              
\usepackage{multirow}
\usepackage{array}
\newcolumntype{P}[1]{>{\centering\arraybackslash}p{#1}}
\newcolumntype{M}[1]{>{\centering\arraybackslash}m{#1}}

\usepackage[pdftex]{graphicx}

\usepackage{hyperref}
\usepackage{etoolbox}
\appto\UrlBreaks{\do\-}

\usepackage{amsmath,amssymb}

\usepackage{changepage}

\usepackage[utf8x]{inputenc}

\usepackage{textcomp,marvosym}

\usepackage{cite}

\usepackage{array}

\newcolumntype{+}{!{\vrule width 2pt}}

\newlength\savedwidth

\title{\LARGE \bf
Polypus: a Big Data Self-Deployable Architecture for Microblogging\\
Text Extraction and Real-Time Sentiment Analysis%
}

\author{%
     Rodrigo Martínez-Castaño, Juan C. Pichel and Pablo Gamallo\\
     Centro de Investigación en Tecnoloxías da Información (CiTIUS)\\
     Universidade de Santiago de Compostela, Spain\\
     \{rodrigo.martinez, juancarlos.pichel, pablo.gamallo\}@usc.es
}

\begin{document}

\maketitle
\thispagestyle{empty}
\pagestyle{empty}

\begin{abstract}

In this paper we propose a new parallel architecture based on Big Data technologies for real-time sentiment analysis on microblogging posts. Polypus is a modular framework that provides the following functionalities: (1) massive text extraction from Twitter, (2) distributed non-relational storage optimized for time range queries, (3) memory-based intermodule buffering, (4) real-time sentiment classification, (5) near real-time keyword sentiment aggregation in time series, (6) a HTTP API to interact with the Polypus cluster and (7) a web interface to analyze results visually. The whole architecture is self-deployable and based on Docker containers.

\end{abstract}

\section{INTRODUCTION}
Sentiment Analysis consists of finding the opinion (e.g., positive, negative, or neutral) from text documents such as movie or product reviews. Opinions about movies, products, etc., can be found in web blogs, social networks, discussion forums, and so on. Companies can improve their products and services on the basis of customer reviews and comments. Recently, many works have stressed the microblogging service Twitter, which generates around 500 million tweets a day\footnote{\url{https://blog.twitter.com/marketing/en_gb/a/en-gb/2015/12-twitter-facts-for-2015.html}}. As Twitter can be seen as a large source of short texts (tweets) containing user opinions, most of these works make sentiment analysis by identifying user attitudes and opinions toward a particular topic or product~\cite{Go2009}.
Perform sentiment analysis on tweets is a hard challenge. 
On the one hand, as in any sentiment analysis framework, we have to deal with human subjectivity. Even humans often disagree on the categorization on the positive or negative sentiment that is supposed to be expressed on a given text~\cite{Villena2013}. 
On the other hand, tweets are too short text to be linguistically analyzed, and it makes the task of finding relevant information (e.g., opinions) much harder. 

Useful conclusions can only be extracted when huge amounts of text or documents are analyzed. However, standard solutions cannot handle gigabytes or terabytes of text data in reasonable time. In this way, professionals demand scalable solutions to boost performance of the sentiment analysis process. 

In this paper we introduce Polypus, a new modular framework based on Big Data technologies designed to perform real-time opinion mining on short texts harvested also in real time. In addition, it stores the retrieved tweets with its metadata and polarity and performs statistics of huge amounts of posts in short times which are visually represented. Our system, deployed in a small cluster, is able to retrieve more than $30$ million tweets a day (with a single target language) and process all of them (supposing matches for all the retrieved tweets) in less than two minutes, whereas more than $1$ million tweets can be processed in around $15$ seconds. The main characteristics 
of Polypus are the following:
\begin{itemize}
\renewcommand\labelitemi{--}
\item A distributed web crawler for Twitter that can retrieve tens of millions of tweets per day was developed, enabling the capacity of building huge corpus of tweets in short times and studying sentiment trends about brands, stocks, products, etc. The goal of the crawler is to demonstrate that the system is capable of processing huge amounts of posts in real time simulating a Twitter \emph{firehose} scenario.
\item The storage solution design, based on HBase~\cite{hbase}, takes into account the problem of hotspotting when using sequential keys. Contemporary posts will be evenly distributed across the cluster at the same time that they are chronologically sorted in every node for efficient scans.
\item The distributed key-value memory-based store Aerospike~\cite{Aerospike} was used as a set of fast communication buffers between some of the modules of Polypus and to avoid storing duplicated tweets by the crawler.
\item The Polypus processing core was implemented as a combination of two Big Data technologies: Storm~\cite{Storm}, for real-time sentiment analysis on the retrieved tweets, and Spark~\cite{Spark}, with the goal of performing near real-time queries on the already classified tweets. Different metrics are returned providing the associated sentiment of the Twitter community for a given keyword (e.g., average polarity, total matches, polarity ratios: positive, negative and neutral).
\item We have developed a HTTP API to interact with the Polypus cluster and a web interface which allows the users to execute and analyze query results in time series, represented with charts and including the most relevant information.
\item The full software infrastructure behind Polypus can be tuned and deployed in a custom cluster in a few minutes with its own deployment mechanism. All the components run inside Docker containers, allowing better resource isolation for the several frameworks and modules at the same time that the infrastructure is guaranteed to work in any docker-compatible Linux cluster.
\item Finally, Polypus is free software\footnote{Available at: \url{https://github.com/polypus-twitter}}.
\end{itemize}

There are many interesting studies in the literature about sentiment analysis on Twitter data that use Big Data processing frameworks, but none of them bring together in the same system all the features of Polypus. Some process tweets in real time with Storm but do not take into account aggregation by keywords~\cite{Agerri201536, karanasou2016} while Polypus considers this kind of processing with arbitrary terms. Other studies use Storm but they are limited to a fixed set of keywords~\cite{wang2012} or Twitter hashtags~\cite{minanovic2014}. In~\cite{rahnama2014}, a real-time system where the users can execute queries for relevant terms is described, but the results refer to the present and thus, historical analysis is not available. With Polypus, queries can be executed for any keyword in any time interval since the system begins its work. In~\cite{wang2012} the cloud-based propietary solution IBM InfoSphere Streams is used to analyze tweets in real time. However, data aggregation happens within their web interface for a limited set of entities. All the main components of the software infrastructure that powers Polypus are open source. Moreover, as previously said, our system is licensed under a free software license. 

Apache Hadoop \cite{trendminer2012, bliu2013, khuc2012} and Apache Spark \cite{assiri2016} are also commonly used for sentiment analysis. In \cite{bharti2016}, sentiment and sarcasm analysis are performed with frequent batch jobs using Hive and Hadoop. Note that Spark can perform micro-batch (Spark Streaming) and batch jobs, whereas Hadoop MapReduce was thought for the single purpose of batch jobs. However, Storm was explicitly designed to perform true real time processing. The distributed processing model of Apache Storm is very suitable for real-time natural language processing (NLP) software since the different steps (pipeline modules) of our sentiment analyzer can be executed concurrently and are permanently processing new data.

Our system combines two technologies for sentiment analysis. On the one hand, we use Storm to classify the retrieved tweets in real time. On the other hand, Spark is used for fast aggregation of polarities in queries with arbitrary keywords. In addition, to the best of our knowledge, there is not any other similar system using Docker containers for effectively resource limitation and fast on-demand deployment.
 
It is worth mentioning that the sentiment classifier included in Polypus performs as well as other state-of-the-art classifiers considering tweets written in different languages and according to the results reported in the following shared tasks: 
TASS~\cite{GamalloTASS2013 ,Villena2013} 
 and SemEval (task 9)~\cite{GamalloSEMEVAL2014, Rosenthal2014}. TASS is an experimental evaluation workshop for sentiment analysis and online reputation analysis focused on Spanish language. SemEval (task 9 or 10) is focused on sentiment analysis in English microblogging.

The paper is structured as follows: Section \ref{sec:background} describes related works and technologies which power Polypus. Section \ref{sec:architecture} explains the functioning of the modules which compose the system. Section \ref{sec:containers} explains the implications of a container-based deployment. In Section \ref{sec:results} experimental results are shown and discussed. Finally, Section \ref{sec:conclusions} contains the main conclusions of the work.
\section{BACKGROUND \& RELATED WORK}
\label{sec:background}
\subsection{Tweet Retrieval}
There are important limitations in the way the tweets can be retrieved via different public Twitter APIs. Due to this fact, it is not possible to retrieve huge amounts of posts in short times (with the exception of paying for the access to the {\it firehose}). 
In this paper, we introduce a web crawler which was executed during a limited period of time with the only purpose of illustrating the functioning of the whole architecture in a real scenario\footnote{\url{http://eur-lex.europa.eu/legal-content/EN/TXT/HTML/?uri=CELEX:31996L0009&from=EN}}\footnote{\url{https://www.boe.es/buscar/pdf/1996/BOE-A-1996-8930-consolidado.pdf}}. 

Twitter provides two public APIs for retrieving tweets: the Streaming API and the HTTP API. The Streaming API serves a limited stream (around 50 tweets per second, from which only around 15 are in English) through a persistent HTTP connection. There are two public streams, both representing around the 1\% of the emitted tweets in every moment: the {\it sample} and the {\it filtered stream}. Both provide a limited access to the Twitter firehose. The filtered stream allows us to set some restrictions, so it is possible to retrieve tweets in concordance with our needs, whereas the sample stream is a random selection. There is a big drawback related to the filters of the filtered stream: it is not possible to filter only by language. 
The HTTP API is synchronous and very limited in the number of calls. 

In order to perform opinion mining, huge amounts of tweets have to be collected. There are different approaches but all of them use the provided public Twitter APIs or the paying access 
to the {\it firehose}~\cite{wang2012}. Several studies perform data extraction through target keywords or through the unfiltered sample stream \cite{dang2016, kumar2016, minanovic2014}. Usually, matching an important event like NBA Playoffs~\cite{baucom2013} or during long intervals\cite{zarrad2014, skuza2015}. Our solution, however, uses a custom crawler in addition to the Streaming API with the aim of feeding our real-time sentiment analyzer with enough data during the experiments. With our approach, similar quantities to the obtained in a full day with the Streaming API in \cite{skuza2015} can be retrieved in around 15 minutes (see Section \ref{subsec:crawler_eval}).

There are some tools developed for gathering and storing tweets. T-Hoarder \cite{hoarder2017} allows the user to collect tweets from the Twitter APIs and store them in files. It also provides a way to perform analytics on the extracted metadata. Other tools store tweets in different relational~\cite{zombie2012,twitterecho2012} or NoSQL databases like MongoDB~\cite{tap2014} and HBase~\cite{minanovic2014}. Some related works use HBase for other tasks as building a lexicon for their classifier\cite{khuc2012}. 
Our approach not only heavily increases the retrieval capacity but also provides an optimal way to store tweets. They can be efficiently retrieved in time ranges and the dataset can grow up horizontally.

\subsection{Sentiment Analysis \& Natural Language Processing}

There are two types of approaches for sentiment analysis: machine learning classification and lexicon-based strategy. 
Machine learning methods use several learning algorithms to determine the sentiment by training on a known dataset. Many of them rely on very basic classifiers, e.g., Naïve Bayes \cite{Kramer2014} or Support Vector Machines\cite{Pang2002}. They are trained on a particular dataset using features such as bag of words or bigrams, and with or without part-of-speech (PoS) tags. 
The lexicon-based technique involves calculating sentiment polarity for a text using dictionaries or lexicons of words. The lexicon entries are annotated with their semantic orientation: polarity (positive, negative or neutral) and strength. 

To deal with short messages such as tweets and SMS, the state-of-the-art systems are based on machine learning techniques using as features polarity lexicons~\cite{giachanou2016}. 
Our strategy also makes use of polarity lexicons to enrich the set of features of the classifier. Most recent approaches to sentiment analysis on short messages and social media are endowed with rich linguistic information such as shallow syntactic structures \cite{Severyn2015} or syntactic dependency trees \cite{Vilares&Alonso2015}. Following the tendency to use knowledge-rich linguistic features, our approach will be provided with shallow syntactic information to detect polarity shifters (e.g., negation markers). 

\subsection{Big Data Technologies}
MapReduce~\cite{Dean2004} is a programming model introduced by Google for processing and generating large data sets on a huge number of computing nodes. A MapReduce program execution is divided into two main phases: \emph{map} and \emph{reduce}. The input and output of a MapReduce computation is a list of key-value pairs. Users only need to focus on implementing map and reduce functions. In the map phase, map workers take as input a list of key-value pairs and generate a set of intermediate output key-value pairs, which are stored in the intermediate storage (i.e., files or in-memory buffers). The reduce function processes each intermediate key and its associated list of values to produce a final dataset of key-value pairs. In this way, map tasks achieve data parallelism, while reduce tasks perform parallel reduction. Currently, several processing frameworks support this programming model. 

Apache Hadoop~\cite{Hadoop} is the most successful open-source implementation of the MapReduce programming model. Hadoop consists of three main layers: a data storage layer (HDFS), a resource manager layer (YARN), and a data processing layer (Hadoop MapReduce Framework). HDFS is a block-oriented distributed file system based on the idea that the most efficient data processing pattern is a write-once, read-many-times pattern. 

We use Apache Spark to aggregate results. It allows us to describe arbitrary workflows with several processing steps, so it is a more flexible model than Hadoop MapReduce. Results can be cached between tasks. It supports several functional programming operations beyond MapReduce tasks. Spark loads the data in memory, without storing intermediate results in disk as Hadoop does. Spark requires a cluster manager and it includes its own one (standalone manager). We used YARN instead as it provides richer resource scheduling capabilities for big clusters. The executors are processes launched for a Spark application and they run units of work called tasks. Jobs are divided into stages (sets of tasks) which are dependent. In addition, another process, called driver, is the responsible of orchestrating the execution of an application and launch the executors.

In order to process the retrieved tweets in real time we use Apache Storm, which is a framework with the aim of processing streaming data in real time. It requires the definition of {\it topologies}, which are computational graphs (workflows) where every node represents individual processing tasks. Edges correspond to the flowing data between nodes, which are the responsible of exchange data using {\it tuples}. Tuples are ordered lists of values, where each value has an assigned name. 
In particular, nodes exchange non delimited sequences of tuples called {\it streams}. Every node listens to one or more streams as input. In the Storm terminology, {\it spouts} are the sources of a stream within a topology, which usually read data from an external source. Finally, {\it bolts} are the consumers of the streams and they perform calculus and transformation tasks on the received data. Bolts can emit none, one or more tuples to the output streams. When a topology starts, it stays in execution waiting for new data to process. 
Storm clusters are composed by two types of nodes: master (Nimbus) and workers (supervisors). As HBase or Hadoop (when installed as a high available cluster), Storm makes use of Apache ZooKeeper\cite{zookeeper} for coordinating the access to the cluster. ZooKeeper is a distributed and highly reliable centralized coordination service which serves other distributed applications.

Tweets in Polypus are stored in Apache HBase, which is a non-relational column-oriented key-value store. Most of these kind of systems have common characteristics. The databases they manage do not require fixed schemas and usually scale up horizontally. Some of them do not fully guarantee ACID and often do not support JOIN-like operations. 
HBase supports replication and uses HDFS for data storage and ZooKeeper for coordination. Tables are composed of column families. Column families are groups of columns that are stored together in disk for the same row. When retrieving data, grouping those columns according to similar access patterns will improve the performance. Every row is identified by a {\it rowkey}. The order in which rows are stored is lexicographical. Due to that, it is important to notice that sequential rowkeys will provoke hotspotting in the node whose region is being written to or read from. This happens because a {\it region} (the most basic element for availability and distribution of Tables) can only be hosted by a single node.

Aerospike is a key-value distributed store which is used in Polypus to store input and output temporary data for some modules. It uses RAM memory or SSD disks for storing data and supports disk persistence when using memory. All the nodes have the same role and support TTL (Time-To-Live) for records, replication, and the definition of UDFs (User Defined Functions). {\it Sets} are analogue to relational tables and belong to {\it namespaces}. We also use MariaDB\cite{MariaDB} to store aggregated results obtained with the Aggregation Module (See Section \ref{subsec:spark_arch}). 

Polypus is deployed using Docker~\cite{docker} containers, which allow us to obtain the benefits of virtualization (isolation, flexibility, portability, agility, etc.) without penalizing the I/O performance considerably. Docker makes use of resource isolation characteristics of the Linux kernel, 
so independent containers can be executed on the same host. Containers supply a virtual environment with their own space of processes and networks. The containers are built with stacked layers. When a container is in execution, a new writable layer is created over a set of read-only layers which define a Docker image. 
Images can be built with a custom scripting language ({\it dockerfiles}) or by saving the state of a running container.
\section{ARCHITECTURE OF THE SYSTEM}
\label{sec:architecture}
%
\begin{figure}[!t]
\centering
\includegraphics[width=0.47\textwidth]{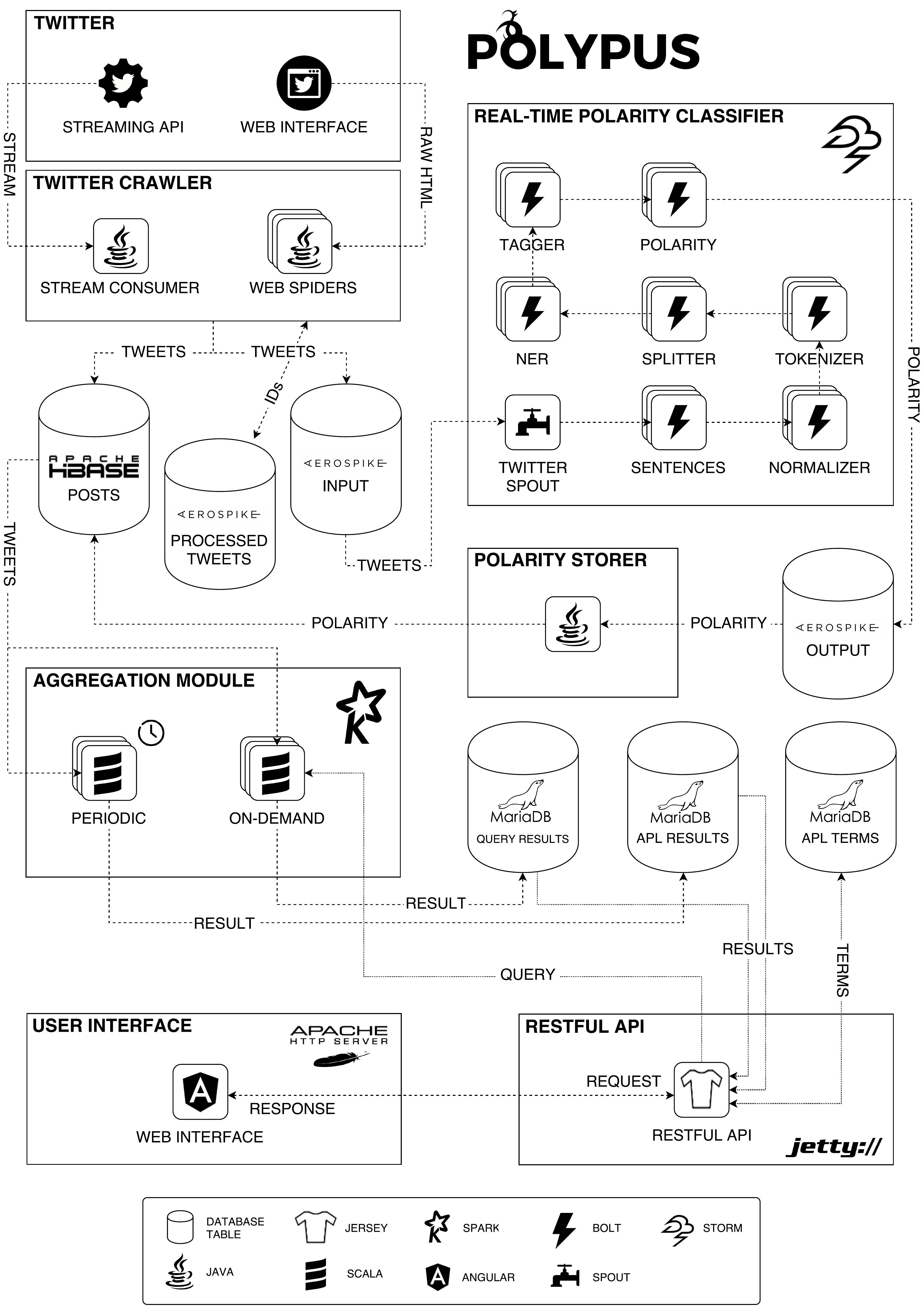}
\vspace{0.2cm}
\caption{Architecture diagram of the Polypus system.}
\label{fig:architecture}
\end{figure}

Our system is composed of several modules which will be explained in the following subsections in the same order in which data flows. From the retrieval of posts to the visualization of results.

\subsection{Polypus4t: A Twitter Crawler}
\label{subsec:crawler_arch}
Polypus4t was developed in Java and consists of two main components. A web crawler which acquires tweets directly from the search web interface and a consumer of the Twitter Streaming API fueled by the Twitter4J\footnote{http://twitter4j.org/} library.

The web crawler retrieves HTML code that the Twitter web client returns for every query built from a list of selected terms in the wanted language or languages. 
These searches are limited to around 20 tweets for every term in the list every 20 seconds. The returned tweets are a sample of the total matching a given term and any user of the web interface will receive the same set of tweets when making the query within the same 20-second window. This method gets good results using frequency lists of certain languages or trending topics, but without a linear relation with the available resources (due to external limitations). When this module is executed, one thread is consuming the Streaming API (optional) while a configurable number of threads (web spiders) are retrieving tweets from the web interface with the list of selected terms. 
Other relevant parameters that can be set are: the number of spiders to launch, the minimum size of the buffer of tweets, the buffer step (allowing to distribute the writes into HBase since the buffer size of each thread is increased progressively with this value), the number of laps (times that every spider checks its assigned terms) and the sleep time (number of seconds that every spider will pause its activity before starting a new lap). It is possible to launch several instances of the module in different nodes. 

In order to avoid repeated tweets, there is an Aerospike set for the task, where the tweet identifiers are stored for two days. Tweets and their metadata are stored in a HBase table. In addition, the content, an internal Polypus post identifier and the language of every tweet are also stored in another Aerospike set: the sentiment classifier input buffer. All the sets are stored in memory for fast reads and writes from the adjacent modules. Before dumping the tweets, new posts are filtered through its Twitter identifier. 

Since HBase stores the rows lexicographically, in order to avoid region hotspotting when writing or scanning posts in time ranges, rowkeys must have a proper design. It is important that the keys have a chronological order so the scans can be performed faster within time intervals. A solution to this problem consists in adding a prefix to the key that will usually cause the assignment of the row to a different region and probably to a different node~\cite{sematext}. The number of prefixes ({\it buckets}) is proportional to the randomness of the data distribution. 

\subsection{Real-Time Sentiment Classifier}
\label{subsec:storm_arch}

Our approach is based on a Naïve Bayes (NB) classifier.
NB combines efficiency (optimal time performance) with reasonable accuracy. The main theoretical drawback of NB methods is that it assumes conditional independence among the linguistic features. If the main features are the tokens extracted from texts, it is evident that they cannot be considered as independent, since words co-occuring in a text are somehow linked by different types of syntactic and semantic dependencies. However, even if NB produces an oversimplified model, its classification decisions are accurate \cite{Manning2008}. To improve the performance of the system, the classifier was enriched with lexicon-based features. Our sentiment analysis system is called \emph{CitiusSentiment}\footnote{Available at: \url{http://gramatica.usc.es/pln/tools/CitiusSentiment.html}}, it is multilingual and supports English and Spanish texts. It was originally implemented in Perl. The modules were rewritten in Java and adapted as Storm bolts.

The Storm topology which process and classifies every tweet makes use of two Aerospike sets that behave as buffers. The first one is the already mentioned input set. The second one is the output set, where the estimated polarity of a post is stored for every custom post identifier. The topology does not use HBase. Due to these buffers there is a very fast exchange of information between the Twitter Crawler, the Storm topology and the Polarity Storer module, which reads the results from the output set and updates the value in the HBase table asynchronously. After updating the HBase table, read results are deleted from the buffer. 

The Storm topology is composed of seven bolts (corresponding to English language) and one spout. More languages does not mean doubling the number of bolt classes, since some modules of the sentiment classifier pipeline are language agnostic. E.g., the Naïve Bayes classifier. The spout is in charge of reading the written posts from the input buffer and distribute them in several streams to the topology, one per language. The bolts perform different tasks in sequence so the polarity can be estimated and written to the output buffer in the last one (Polarity bolt):
\begin{itemize}
\item Sentences: it splits the post text in sentences.
\item Normalizer: swaps some elements like abbreviations or emoticons, among others, for semantic tags.
\item Tokenizer: every sentence is transformed in a token sequence.
\item Splitter: transforms the composed words in contractions. E.g., {\it don't = do + not, we'll = we + will}.
\item NER: it recognizes named entities which can contain several words. For instance: Santiago de Compostela.
\item Tagger: the PoS Tagger performs a morphosyntactic tagging. E.g., {\it Proper noun, singular (NNP); Verb, 3rd person singular present (VBZ)}.
\item Polarity: it uses a polarized lexicon and a Naïve Bayes classifier trained with tweets. The result is an integer value: -1 (negative), 0 (objective) or 1 (positive).
\end{itemize}

The topology, as mentioned, is a simple sequence, so each bolt emits tuples in a custom stream read only by an instance of the next bolt in the pipeline. The first bolt, ``Sentences'', reads directly from the spout, whereas the last one, ``Polarity'', does not emit any tuple. However, it writes the result in the output buffer, associated with the post identifier.

\subsection{Aggregation Module}
\label{subsec:spark_arch}

The Aggregation Module is implemented in Scala and runs on Spark. There are two modes of execution for this module. The first is periodic and automated. There is a list of keywords, called APL (Automated Processing List) which contains relevant keywords that worth to query every few minutes. This allows the user to observe the evolution of those keywords without long waits for heavy on-demand batch jobs. The second mode of execution is driven by custom queries performed in real time with arbitrary keywords. This execution method implies waiting times from seconds to minutes for reasonable time-bounded queries. 
Parallel reads are performed for every prefix (buckets), avoiding the slowdown of the distributed rowkeys in different regions within the same time interval. As the rowkeys are chronologically sorted in every region, it is easy to perform a time range scan with starting and stopping dates and times. The posts fitting the defined time interval which contain the keywords of the query will be filtered. The aggregation operations are made globally in the selected time interval but also in windows within the interval. The window size is defined by the user. In a map function, for each matching post, a 4-tuple is emitted as a value and the matching window (a timestamp), as the key. Once all the tuples are aggregated (sum), their values represent the average keyword polarity (taking into account positive and negative but also the neutral results which soft the aggregated polarity), the total number of positive posts and, lastly, the number of negative ones. The normalized average polarity (AP) is calculated as follows, where ${p_i}$ is the polarity of every post and ${m_k}$ is the number of matches for a given keyword: 
\[ AP = (\sum\limits_{i=1}^{m_k} (p_i) / m_k + 1)/2 \]

From the aggregated values, the total number of neutral matches and the ratio of positive and negative posts are obtained. All of these values are stored in a MariaDB table which varies with the execution mode (automated or on-demand) as it can be observed in Figure \ref{fig:architecture}.

\subsection{HTTP API}
\label{subsec:api_arch}

The HTTP API was implemented in Java with the Jersey\footnote{\url{https://jersey.java.net/}} framework. It is the interface of the Polypus back-end. Directly or through a web interface, it allows the user to perform all the available operations. Currently, the following functionalities are implemented:
\begin{itemize}
\item Obtaining the keywords of the APL (Automated Processing List).
\item Searching for keywords of the APL by name.
\item Obtaining a custom number of trending keywords of the APL.
\item Obtaining the evolution of a keyword of the APL in a time interval.
\item Obtaining the next measure of a keyword of the APL after a date and time.
\item Deleting a keyword from the APL.
\item Adding a keyword to the APL.
\item Obtaining the results of a custom query.
\item Searching for completed custom queries.
\item Launching a custom query. 
\end{itemize}

\subsection{Graphic User Interface}
\label{subsec:gui_arch}

\begin{figure}[t]
\centering
\setlength{\fboxsep}{0pt}%
\setlength{\fboxrule}{0pt}
\fbox{\includegraphics[width=0.47\textwidth]{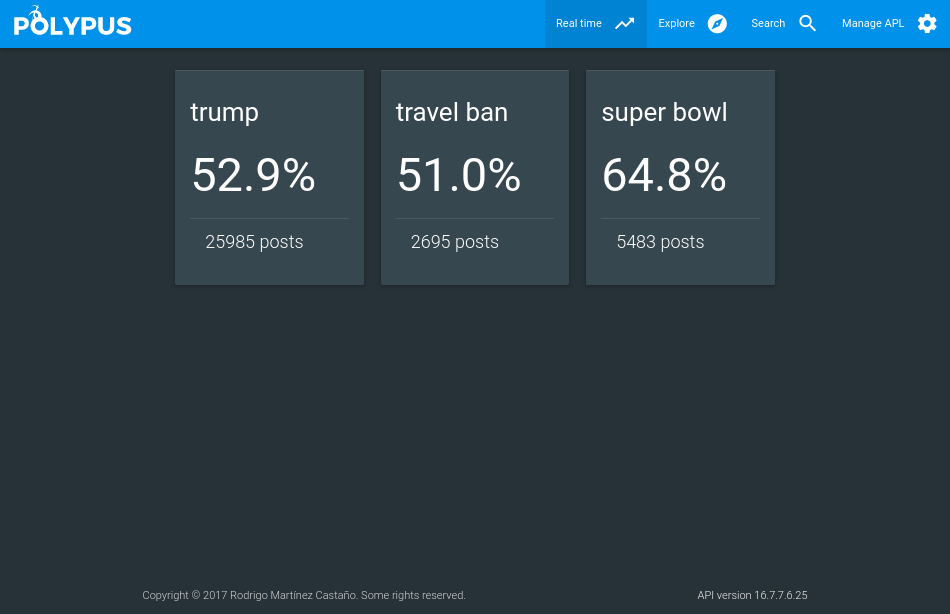}}
\vspace{0.2cm}
\caption{Real-time information for the most relevant keywords of the APL in the last hour.}
\label{fig:realtime}
\end{figure}

\begin{figure}[t]
\centering
\setlength{\fboxsep}{0pt}%
\setlength{\fboxrule}{0pt}
\fbox{\includegraphics[width=0.47\textwidth]{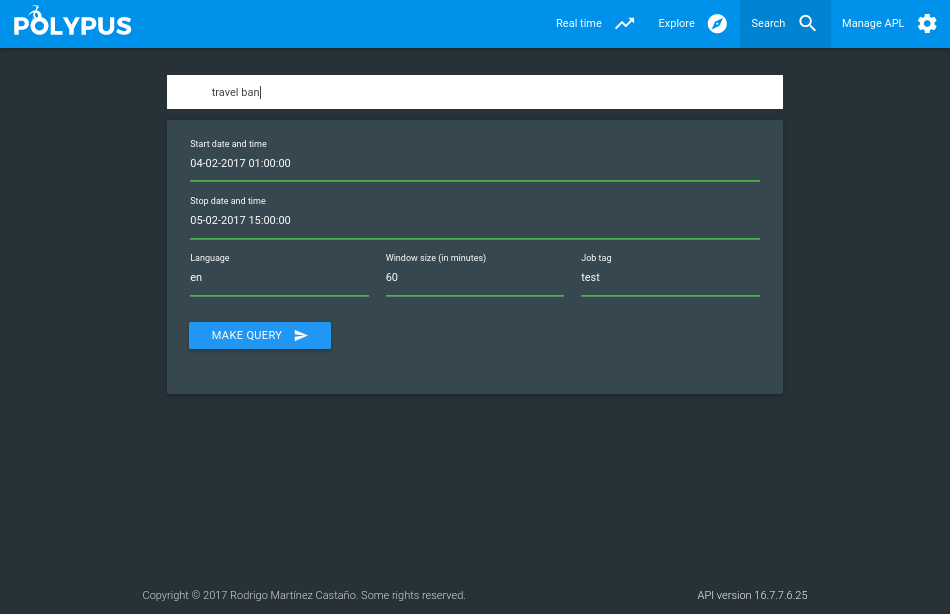}}
\vspace{0.2cm}
\caption{Search view for querying the keyword ``travel ban'' with 60-minute windows within the 37-hour interval of the main experiment.}
\label{fig:search1}
\end{figure}

\begin{figure}[!t]
\centering
\setlength{\fboxsep}{0pt}%
\setlength{\fboxrule}{0pt}
\fbox{\includegraphics[width=0.47\textwidth]{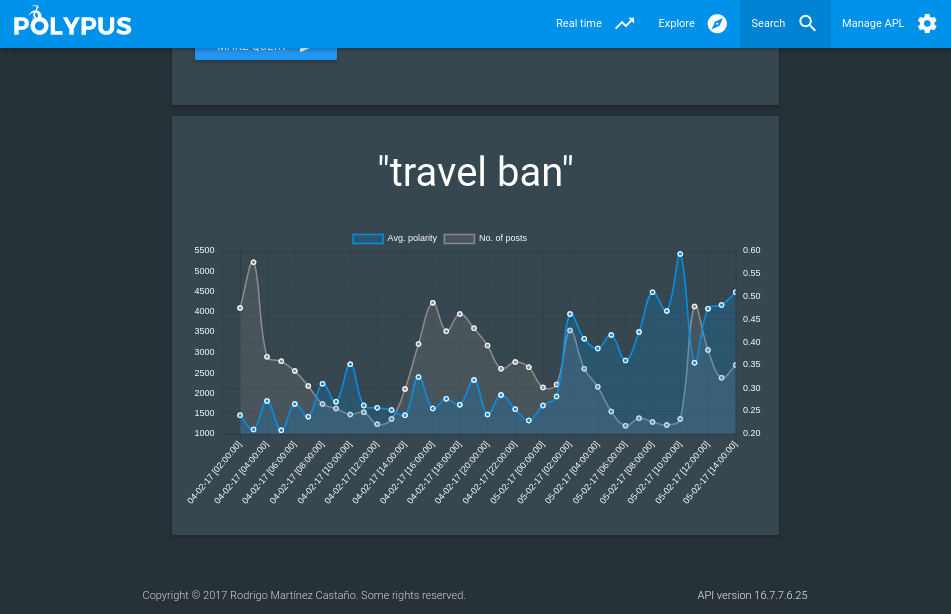}}
\vspace{0.2cm}
\caption{Result of the execution of the query \ref{fig:search1}. Avg. polarity and matching tweets are shown per 60-minute interval.}
\label{fig:result}
\end{figure}

\begin{figure}[!t]
\centering
\setlength{\fboxsep}{0pt}%
\setlength{\fboxrule}{0pt}
\fbox{\includegraphics[width=0.47\textwidth]{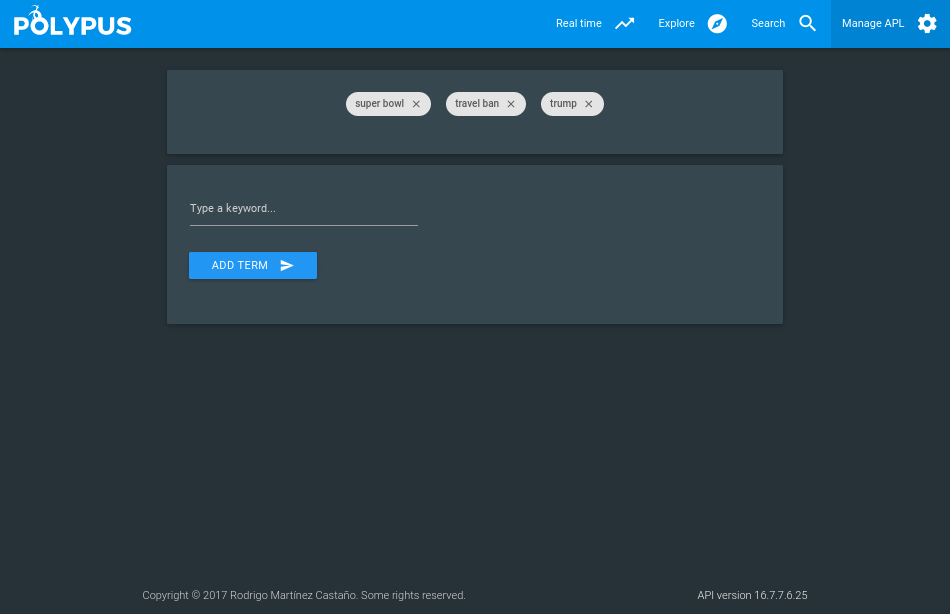}}
\vspace{0.2cm}
\caption{Management of the APL list. Keywords can be added or deleted.}
\label{fig:manage}
\end{figure}

A web interface was implemented to interact with the system. This interface manages some of the implemented functionality of the HTTP API and has four views: {\it Real time, Explore, Search} and {\it Manage APL}. The {\it Real time} view (Figure \ref{fig:realtime}) shows the most active keywords of the APL in real time (those with more retrieved tweets). Through the {\it Search view} (Figure \ref{fig:search1}), users can make custom queries on the full set of classified tweets, specifying mainly: a keyword, start and stop points, language and window size (size of each chunk within the given interval). When the query is completed, the polarity and matches evolution for the keyword is shown in an area chart (Figure \ref{fig:result}). The {\it Explore} section is similar to the {\it Search} view but queries are not executed with the Aggregation Module. Queries can only be formulated with the keywords of the APL. For every keyword, the polarity is aggregated periodically, so the task of aggregating already compacted values is fast. The {\it Manage APL} view is shown in Figure \ref{fig:manage}. It allows the user to modify the existing keywords of the APL.

%
\section{DEPLOYMENT WITH CONTAINERS}
\label{sec:containers}
The full architecture of Polypus is composed by several distributed computation frameworks and databases. When the available resources are limited, the way they are distributed is important. Most of the frameworks do not allow the user to limit physical cores or threads. YARN configurations cover the number of vCores (but this does not limit the CPU usage). The only configurable resource in JVM based frameworks seems to be the memory. A full manual setup of all the Polypus dependencies may take hours and this task have to be repeated when moving to a different cluster. 

Docker allows the user to limit the assigned resources of a container: e.g., the number of cores, RAM memory. Several Docker images were made for each component and scripts to launch and orchestrate the different services. The scripts 
were designed for AWS (Amazon Web Services) EC2 machines, but can be easily adapted to other cloud platforms or local clusters with a few changes in the configuration files. Each service has a custom configuration file for easy tuning the most important parameters (resources, hosts, service own configurations, etc.). A main configuration file defines the full architecture to be deployed (what containers and what services will run on each host).

Considering this deployment system, the full architecture is running on a cluster in about 2 minutes.
\section{EXPERIMENTAL RESULTS} 
\label{sec:results}
In order to test Polypus, we set up the architecture with AWS EC2 virtual machines running Amazon Linux AMI. Amazon gives their users the possibility of running a wide variety of virtual machines in their EC2 infrastructure. In our case, the cluster was created with 3 and 7 nodes of the \texttt{c4.4xlarge} instance type. The used EC2 instances had the following characteristics:
\begin{itemize}
\item CPU: Intel Xeon E5-2666 v3 (Haswell microarchitecture)
\item vCores per node: 16 (8 physical dedicated cores)
\item RAM Memory per node: 30 GiB
\item Disk: each node has five 50 GiB {\it SSD General Purpose} disks. Four of them are reserved for its use in HDFS.
\end{itemize}

According to the specifications provided by Amazon, the \texttt{c4.4xlarge} instances have a dedicated bandwidth between 500Mbps and 4Gbps for EBS storage and a 10Gbps network. 

\begin{figure}[!t]
\centering
\includegraphics[width=0.4\textwidth]{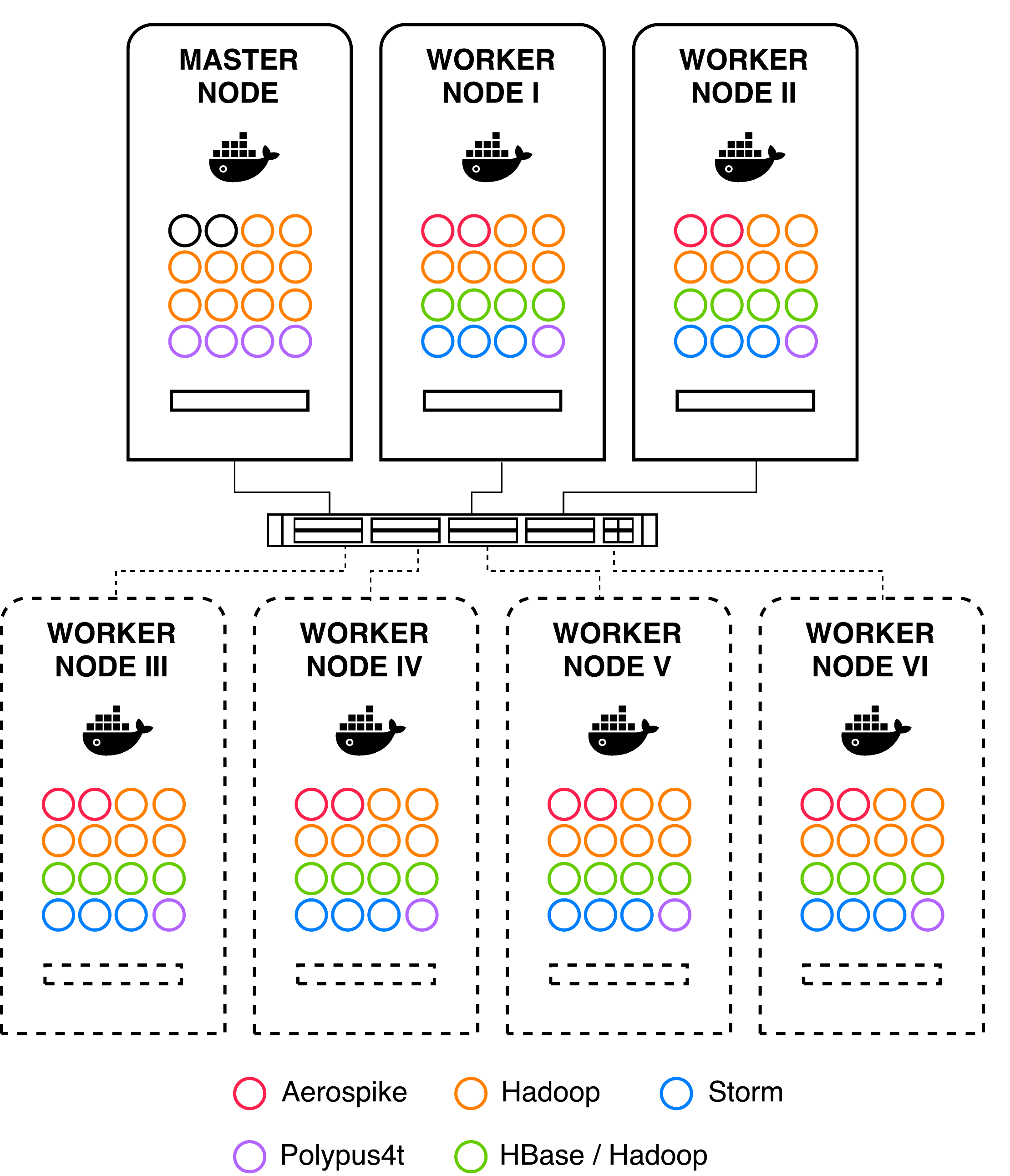}
\vspace{0.2cm}
\caption{AWS EC2 cluster used in the performance tests and core assignment (3-node and 7-node clusters).}
\label{fig:nodes}
\end{figure}

Two-size clusters were used in the tests (see Figure \ref{fig:nodes}): 3-node and 7-node. In both, a {\it master} node contains all the master services. Mainly: {\it Nimbus} for Storm, {\it NameNode} for HDFS, {\it ResourceManager} for YARN, {\it HMaster} for HBase and one of the three servers of ZooKeeper. Also the HTTP API, the Polarity Storer module, the database which stores the Aggregation Module results and an instance of the crawler Polypus4t. A {\it nodemanager} is configured in the master so the ApplicationMaster of the Spark jobs does not fill the assigned resources of a worker. The worker nodes contain the worker processes, Aerospike peers and the remaining ZooKeeper servers. The slave processes are the {\it regionservers} for HBase, {\it nodemanagers} for YARN, {\it datanodes} for HDFS, {\it supervisors} for Storm and a crawler instance. 
\begin{table}[!t]
\renewcommand{\arraystretch}{1.3}
\caption{Distribution of resources: vCores / memory}
\label{resources}
\vspace{0.2cm}
\centering
\footnotesize
\begin{tabular}{|c||c|c|}
\hline
\bfseries System & \bfseries Master & \bfseries Worker\\
\hline\hline
Aerospike & - & 2 / 3 GiB \\ \hline
Hadoop & 10 / 17 GiB & 10* / 17 GiB \\ \hline
HBase & unrestricted & 4 / 4 GiB \\ \hline
Storm & unrestricted & 3 / 2 GiB \\ \hline
Polypus4t & 4 / 3 GiB & 1 / 2 GiB \\
\hline

\end{tabular}
\\
\vspace{1em}
*4 vCores are shared with HBase
\end{table}

\subsection{Twitter Crawler Evaluation}
\label{subsec:crawler_eval}

\begin{figure}[!t]
\centering
\includegraphics[width=0.47\textwidth]{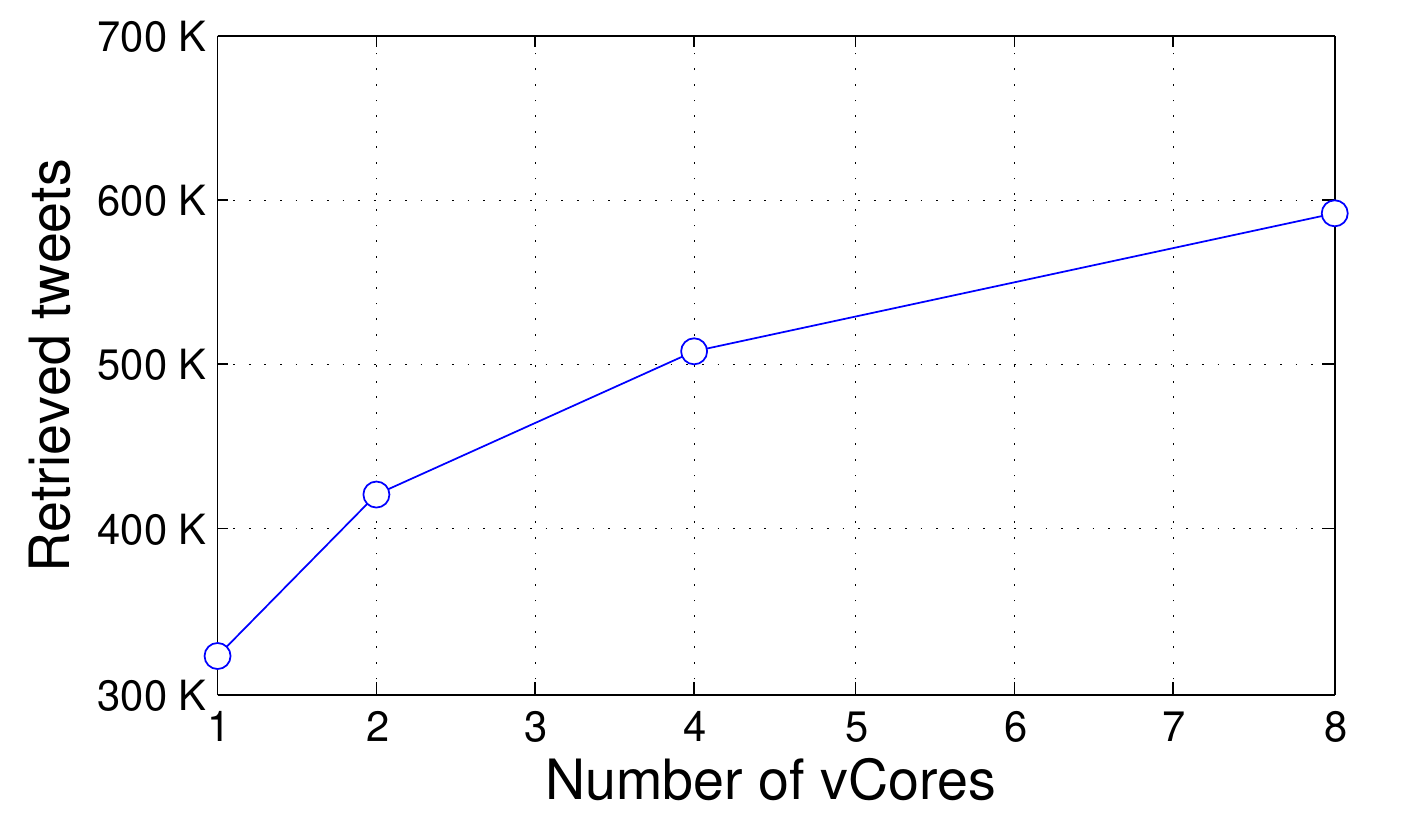}
\vspace{0.2cm}
\caption{Retrieval capacity in a single node with {\it n} cores and 32{\it n} threads per core.}
\label{retrieval-cores}
\end{figure}

\begin{figure}[!t]
\centering
\hspace{20cm}
\includegraphics[width=0.47\textwidth]{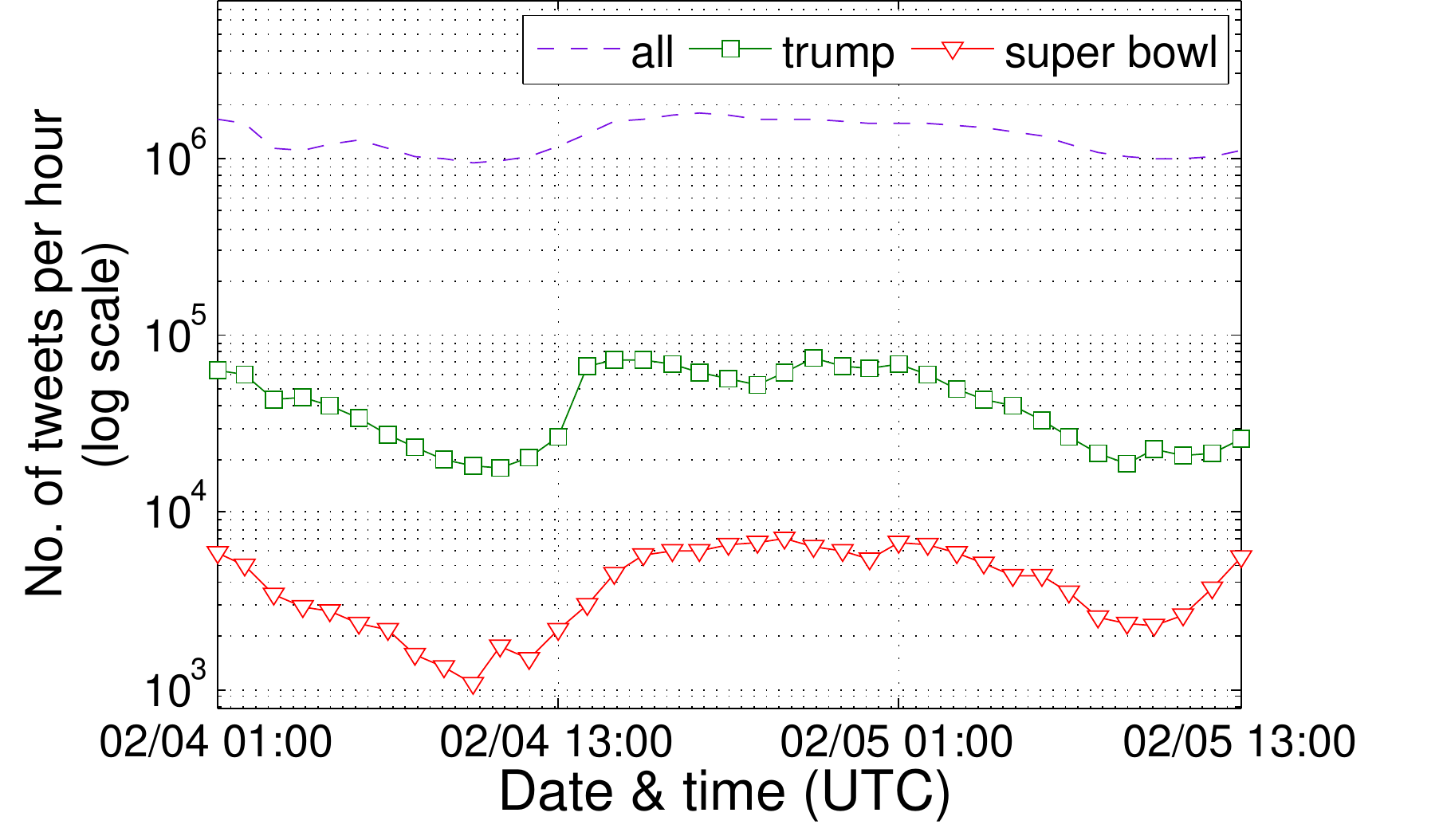}
\vspace{0.17cm}
\caption{Number of collected tweets per hour for the keywords ``super bowl'' and ``trump'' over the 37-hour experiment.}
\label{retrieval-37h}
\end{figure}

In order to test the capabilities of this module, the target terms were two combined lists of the most frequent words for the English language. The terms were extracted from a $6,000$ words list (insightin.com\footnote{Available at: \url{http://www.insightin.com/esl/}}) and a $10,000$ words list (collected by Eric Price\footnote{Available at: \url{http://www.mit.edu/~ecprice/wordlist.10000}}). The result was a new list of $11,329$ words.

Some tests were made in order to determine the most equilibrated configuration and to obtain the retrieval capacity of the crawler. The Streaming API was not used in these tests. In 15-minute intervals, the first step was to obtain the optimum number of threads per core, which resulted to be 32 (see Table \ref{tweets-1-core}). Additional tests were performed considering 2, 4 and 8 cores with 32 threads per core. It can be observed that the scalability is quiet limited in a single node (Figure \ref{retrieval-cores}). 
\begin{table}[!t]
\renewcommand{\arraystretch}{1.3}
\caption{Number of threads and retrieval capacity of the crawler in a single node with 1 available core (15-minute experiments).}
\label{tweets-1-core}
\vspace{0.2cm}
\centering
\footnotesize
\begin{tabular}{|c|c|c|c|}
\hline
\bfseries Threads (1 core) & \bfseries Tweets & \bfseries Tweets per second\\
\hline\hline
8 & $152,353$ & $169.3$ \\ \hline
16 & $227,114$ & $252.4$ \\ \hline
\bf 32 & $\bf265,801$ &  $\bf295.3$ \\ \hline
64 & $264,932$ & $294.4$ \\ \hline
128 & $262,238$ & $291.4$ \\
\hline
\end{tabular}
\end{table}

Note that these results may not be directly comparable as they have been obtained from consecutive executions and the retrieval capacity has big variations depending on the number of published tweets matching the target terms at a given time.

The crawler was also tested running in distributed mode with two configurations on both clusters: 8 cores in the master and 1 for each worker, and 1 core for all the nodes. The maximum overall performance results are shown in Table \ref{2cluster-crawler}. As it was expected, more cores implies a greater retrieval capacity with a limited scalability. Increasing the number of nodes with a single assigned core from 3 to 7 nodes improves about $20\%$ the retrieval capacity. Something similar happens when incrementing the number of cores to 8 in one of the nodes; improvements about $16\%$ and $11\%$ were observed for the 3-node and 7-node cluster respectively.

\begin{table*}[!t]
\renewcommand{\arraystretch}{1.3}
\caption{Maximum crawler performance for different distributed configurations (15-minute experiments).}
\label{2cluster-crawler}
\vspace{0.2cm}
\centering
\footnotesize
\begin{tabular}{|c|c|c|c|c|}
\hline
\bfseries Cluster size & \textbf{Config.} & \bfseries Nodes, vCores & \bfseries Tweets & \bfseries Tweets per second\\
\hline\hline
\multirow{2}{*}{3-node} 
& a & (3,1) & $500,973$ & $556.6$ \\\cline{2-5}
& b & (1,8), (2,1) & $578,933$ & $643.3$ \\ \hline\hline
\multirow{2}{*}{\bf 7-node} & c & (7,1) & $601,879$ & $668.8$ \\\cline{2-5}
& d & \bf (1,8), (6,1) & $\bf666,579$ & $\bf740.6$ \\ 
\hline
\end{tabular}
\end{table*}

A long-term test in February 2017 (during 37 hours) was performed where the crawler and the real-time sentiment classifier ran unattended. 4 cores were assigned to the master node and 1 for the workers. A total number of $49,650,935$ tweets were retrieved in this period for the only purpose of performing these experiments. In order to illustrate the results, Figure \ref{retrieval-37h} shows the number of retrieved tweets per hour considering two keywords in the test.

\subsection{Real-Time Sentiment Classifier Evaluation}
\label{subsec:storm_eval}

\begin{table}[!t]
\renewcommand{\arraystretch}{1.3}
\caption{Percentage of processed tweets after 15 minutes of execution for the 3-node cluster with 2 Storm worker nodes.}
\label{hints}
\vspace{0.2cm}
\centering
\footnotesize
\begin{tabular}{|c|c|c|}
\hline
\multicolumn{1}{|p{1.9cm}|}{\centering \bfseries Cores per \\ worker node} 
& \multicolumn{1}{|p{1.9cm}|}{\centering \bfseries Parallelism \\ hints}
& \multicolumn{1}{|p{2.1cm}|}{\centering \bfseries Percentage of \\ completion*}
\\\hline\hline

\multirow{3}{*}{2} & 1-1-1-1-1-1-1 & 17.3\% \\\cline{2-3} 
&  1-1-1-1-2-4-1 & 46.9\% \\\cline{2-3} 
&  1-1-2-1-4-16-4 & 37.6\% \\\cline{2-3}
\hline\hline
\multirow{9}{*}{\bf 3} & 1-1-1-1-1-1-1 & 18.8\% \\\cline{2-3} 
&  1-1-2-1-4-16-4 & 61.7\% \\\cline{2-3}
&  1-1-2-1-4-32-4 & 55.5\% \\\cline{2-3} 
&  2-2-2-2-2-4-2 & 69.6\% \\\cline{2-3} 
&  2-2-1-1-1-4-1 & 67.0\% \\\cline{2-3} 
&  1-1-1-1-2-8-1 & 46.4\% \\\cline{2-3} 
&  \bf 1-1-1-1-2-6-2 & \bf 76.3\% \\\cline{2-3}
&  1-1-1-1-2-4-1 & 71.5\% \\\cline{2-3}
&  1-1-1-1-2-6-1 & 54.9\% \\
\hline
\end{tabular}
\\
\vspace{1em}
*With respect to the maximum collected tweets in the tests of $529,820$ units in 15 minutes ($588.7$ tweets per second). The parallelism hints correspond with the different bolts in the following order: {\it Sentences, Normalizer, Tokens, Splits, NER, Tagger} and {\it Polarity}.
\end{table}
\begin{table*}[!t]
\renewcommand{\arraystretch}{1.3}
\caption{Avg. execution time (in minutes) for different configurations when processing the full dataset (around 50M tweets)}
\label{jobs-at-a-time}
\vspace{0.1cm}
\centering
\footnotesize
\begin{tabular}{|c|c|c|c|c|c|c|}
\hline
\multicolumn{1}{|p{1.8cm}|}{\centering \bfseries Cluster \\ size}
& \multicolumn{1}{|p{1.8cm}|}{\centering \bfseries Executors \\ per query}
& \multicolumn{1}{|p{1.8cm}|}{\centering \bfseries Cores per \\ executor}
& \multicolumn{1}{|p{1.8cm}|}{\centering \bfseries Memory per \\ executor}
& \multicolumn{1}{|p{1.8cm}|}{\centering \bfseries Parallel \\ queries}
& \multicolumn{1}{|p{1.8cm}|}{\centering \bfseries Time for a\\ single query}
& \multicolumn{1}{|p{1.8cm}|}{\centering \bfseries Time for \\ 18 queries}
\\\hline\hline
\multirow{3}{*}{3-node} & 
2 & 6 & 14 GiB & 1 & \bf 3.8 & 68.0 \\ \cline{2-7}
& 2 & 3 & 7 GiB & 2 & 6.7 & 60.5 \\ \cline{2-7}
& 2 & 2 & 4 GiB & \bf 3 & 9.1  & \bf 54.4 \\ \hline\hline
\multirow{6}{*}{7-node} & 
6 & 6 & 14 GiB & 1 & \bf 1.9 & 33.7 \\ \cline{2-7}
& 6 & 3 & 7 GiB & 2 & 2.7 & 25.1 \\ \cline{2-7}
& 6 & 2 & 4 GiB & 3 & 3.7 & 22.6 \\ \cline{2-7}
& 2 & 6 & 14 GiB & 3 & 3.8 & 23.8 \\ \cline{2-7}
& 2 & 3 & 7 GiB & 6 & 6.7 & 22.8 \\ \cline{2-7}
& 2 & 2 & 4 GiB & \bf 9 & 9.2 & \bf 20.4 \\ 
\hline
\end{tabular}
\end{table*}

Since our hardware resources are limited, we decided to maximize the assigned resources to the Aggregation Module considering that it has to perform heavy queries in short times. Due to this fact, when the Storm topology begins its work, some bolts delay their start. As a consequence our sentiment classifier has a small warm-up period. After that period all the incoming tweets are processed in real time.

We performed several tests to select an optimal value for the parallelism hints of each Storm bolt. The parallelism hint refers to the initial number of threads of a bolt or spout. 
Table \ref{hints} shows, as example, the percentage of analyzed posts for different configurations in 15-minute tests for the 3-node cluster using 2 Storm worker nodes. This percentage was calculated with respect to the maximum number of tweets collected during these experiments ($588.7$ tweets per second). Note that the crawler uses configuration $b$ of Table \ref{2cluster-crawler}. As it can be observed, 2 Storm worker nodes with the assigned resources cannot handle all the retrieved tweets after 15 minutes. Considering the best parallelism configuration, more than 76\% of the tweets are classified. If the system keeps running, the warm-up period required to process the tweets in real time is 40 minutes (using 3 cores per worker node).

In a different experiment we decreased the resources assigned to the crawler in such a way that the average tweets per second collected during the test drops to $382.8$ (configuration $a$ of Table \ref{2cluster-crawler}). In this case, considering the best parallelism hint, 2 Storm worker nodes are enough to reach real time processing after only 8 minutes for the 3-node cluster.

However, the 7-node cluster cannot reach real-time processing with 2 worker nodes because of a higher tweet retrieval capacity that achieves more than 650 tweets per second (see configurations $c$ and $d$ in Table \ref{2cluster-crawler}). Nevertheless, with an additional worker node, the Storm topology classifies tweets as soon as they arrive to the system also after 8 minutes for all the experiments carried out in this section.

\begin{figure}[!t]
\centering
\includegraphics[width=.47\textwidth]{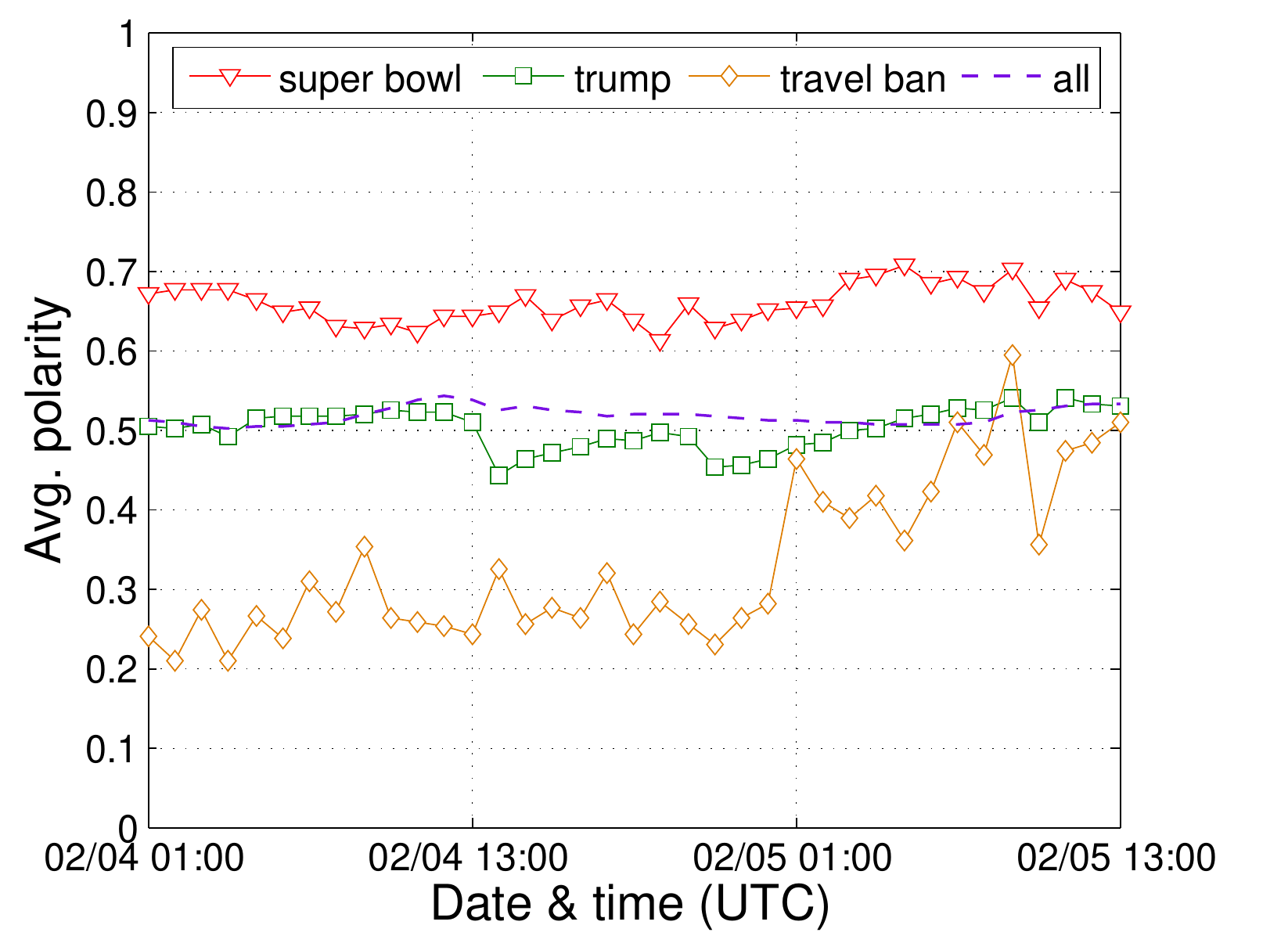}
\caption{Average polarity evolution for the keywords ``super bowl'', ``trump'' and ``travel ban'' over the 37-hour experiment.}
\label{polarity-37}
\vspace{-.1cm}
\end{figure}

In the 37-hour experiment, the polarity evolution for some keywords can be observed in Figure \ref{polarity-37}. The keyword ``travel ban'' increases its positivity after 1:00 a.m. (UTC) of the 5th of February. This can be explained with the suspension of ``all actions'' related to the implementing travel order of the POTUS\footnote{\url{https://www.dhs.gov/news/2017/02/04/dhs-statement-compliance-recent-court-order}}.

\subsection{Aggregation Module Evaluation}
\label{subsec:spark_eval}

\begin{figure}[t]
\centering
\includegraphics[width=.45\textwidth]{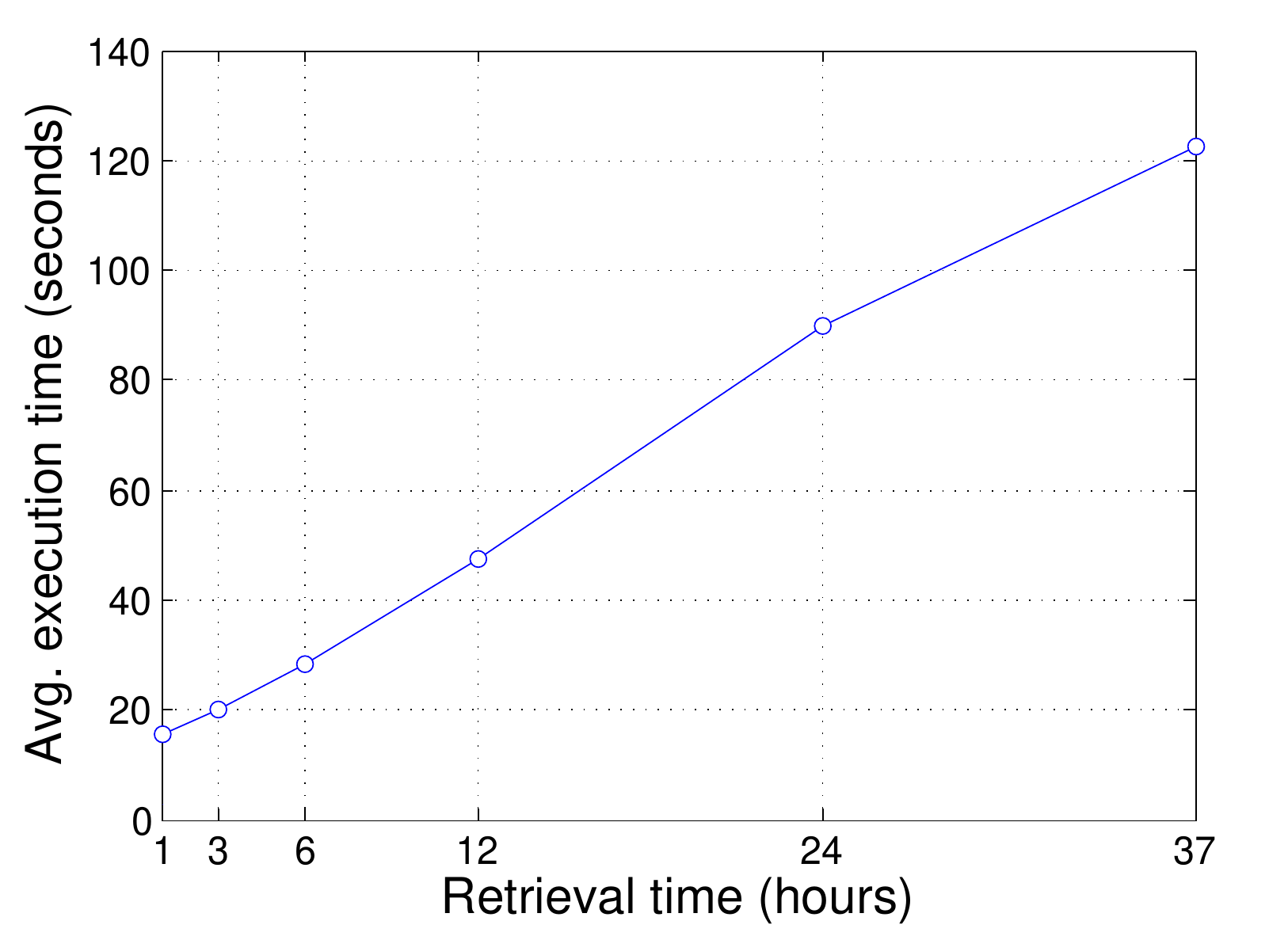}
\vspace{0.2cm}
\caption{Average execution time of the Aggregation Module with 5 cores and 14 GiB of RAM per executor, making queries on the set of collected and analyzed tweets retrieved {\it n} hours backwards. 1.35M tweets collected per hour in average.}
\label{hours}
\end{figure}

\begin{figure}[t]
\centering
\includegraphics[width=.47\textwidth]{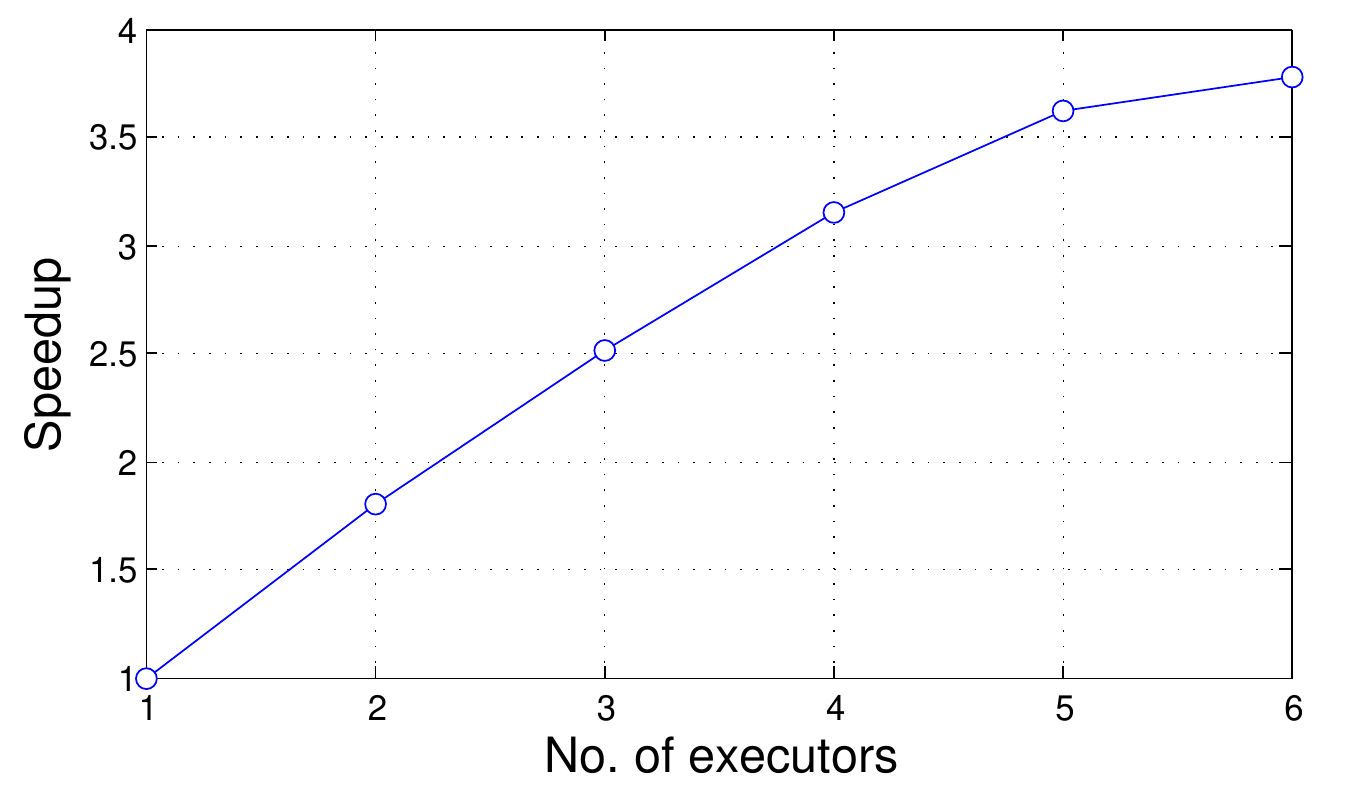}
\caption{Average speedup of the Aggregation Module with 5 cores and 14 GiB of RAM per executor for {\it n} executors when querying the full dataset.}
\label{speedup}
\end{figure}

Table \ref{jobs-at-a-time} shows the average execution times for different combinations of Spark executors and cores and memory assigned to every executor. 
The parallel queries column refers to the number of queries that can be executed at the same time with the available resources. 

As it can be observed, the Aggregation Module performs better with lower number of executors and lower number of assigned cores and memory per executor if there are many jobs running at a time. For non-concurrent jobs, a greater number of cores and memory is always better for the tested configurations. In Figure \ref{hours}, the execution time of processing backwards the last {\it n} hours is shown; there is a linear relation between the retrieval time and the execution time. Figure \ref{speedup} confirms the good behavior of the Aggregation Module in terms of speedup. The small decrease in the speedup using 6 executors is caused by the limited resources assigned to HBase and HDFS.
\section{CONCLUSIONS}
\label{sec:conclusions}

In this paper we present Polypus, a new parallel architecture based on Big Data technologies for real-time sentiment analysis on microblogging posts. Polypus allows the users to create huge datasets of tweets in short times and analyze polarity trends about arbitrary keywords (brands, stocks, products, etc.) in near real time. The architecture is easily deployable with its own scripts within Docker containers.

Polypus currently works with Twitter as data source and with an English sentiment analyzer. However, it can easily incorporate new data sources and languages, considering the same multilingual sentiment classifier or any other. It can estimate the polarity of the retrieved posts in real time through the execution of a Storm topology. 

Aerospike memory based buffers are used to reduce the latency between the modules and to avoid storing repeated posts retrieved by the crawler. 

Spark is used to aggregate results taking into account filters like time ranges and keywords. As aggregated results, we obtain the average polarity, the total number of matches and other relevant measurements like positive, neutral and negative ratios for a given keyword in a time interval. The platform can run automated jobs periodically for a list of target keywords, enabling fast queries for the preprocessed keywords. For custom queries, considering the 7-node cluster, $50$ million tweets (retrieved in a day and a half) can be processed in two minutes, whereas more than $1$ million tweets can be processed in around $15$ seconds. 


A HTTP API makes possible the interaction with the system as a whole, and a web interface allows the user to graphically interact with it and watch the results in charts. 

In addition, the computing resources of the cluster can be limited for the different components, both with centralized configurations and with the isolation capabilities of Docker. This enables fast and automated deployments and cleanliness when stopping using Polypus in a given cluster: either bare metal or cloud based. 

Finally, we must highlight that other existent systems like T-Hoarder do not offer this kind of versatility in the deployment phase and are very limited in their crawling capacity. Others, which use Big Data processing frameworks, do not cover further aspects like querying sentiment associated with arbitrary keywords in near real-time and limit their work to the real-time analysis for individual texts. To the best of our knowledge, there is not any study that covers the problem thinking in horizontal scalability, combining Big Data frameworks and covering all the steps explained in this paper: from the Twitter crawler and polarity assignment in real time to the arbitrary queries in near real-time, allowing the user to analyze the evolution of a given keyword in user-definable time ranges.
\\
\section*{ACKNOWLEDGMENT}
This work has been supported by MINECO (TIN2014-54565-JIN and TIN2016-76373-P), Xunta de Galicia (ED431G/08), European Regional Development Fund and AWS Cloud Credits for Research program. There
was no additional external funding received for this study.
\newpage

\bibliographystyle{ieee_custom}
\bibliography{paper.bib}

\begin{thebibliography}{10}

\bibitem{Aerospike}
{Aerospike}.
\newblock \url{https://www.aerospike.com}.
\newblock [Online; accessed July, 2017].

\bibitem{Agerri201536}
R.~Agerri, X.~Artola, Z.~Beloki, G.~Rigau, and A.~Soroa.
\newblock Big data for Natural Language Processing: A streaming approach.
\newblock {\em Knowledge-Based Systems}, 79:36 -- 42, 2015.

\bibitem{Hadoop}
{Apache Hadoop}.
\newblock \url{https://hadoop.apache.org}.
\newblock [Online; accessed July, 2017].

\bibitem{hbase}
{Apache HBase}.
\newblock \url{https://hbase.apache.org/}.
\newblock [Online; accessed July, 2017].

\bibitem{Spark}
{Apache Spark}.
\newblock \url{https://spark.apache.org}.
\newblock [Online; accessed July, 2017].

\bibitem{Storm}
{Apache Storm}.
\newblock \url{https://storm.apache.org}.
\newblock [Online; accessed July, 2017].

\bibitem{zookeeper}
{Apache Zookeeper}.
\newblock \url{https://zookeeper.apache.org/}.
\newblock [Online; accessed July, 2017].

\bibitem{assiri2016}
A.~Assiri, A.~Emam, and H.~Al-dossari.
\newblock Real-time sentiment analysis of Saudi dialect tweets using SPARK.
\newblock In {\em Proc. of the IEEE Int. Conf. on Big Data (Big Data)}, pages
  3947--3950, Dec 2016.

\bibitem{sematext}
A.~Baranau.
\newblock HBaseWD: Avoid RegionServer Hotspotting Despite Sequential Keys.
\newblock
  \url{https://sematext.com/blog/hbasewd-avoid-regionserver-hotspotting-despite-writing-records-with-sequential-keys},
  2012.
\newblock [Online; accessed July, 2017].

\bibitem{baucom2013}
E.~Baucom, A.~Sanjari, X.~Liu, and M.~Chen.
\newblock Mirroring the Real World in Social Media: Twitter, Geolocation, and
  Sentiment Analysis.
\newblock In {\em Proc. of the Int. Workshop on Mining Unstructured Big Data
  Using Natural Language Processing}, pages 61--68, 2013.

\bibitem{bharti2016}
S.~Bharti, B.~Vachha, R.~Pradhan, K.~Babu, and S.~Jena.
\newblock Sarcastic sentiment detection in tweets streamed in real time: a big
  data approach.
\newblock {\em Digital Communications and Networks}, 2(3):108 -- 121, 2016.

\bibitem{zombie2012}
A.~Black, C.~Mascaro, M.~Gallagher, and S.~P. Goggins.
\newblock Twitter Zombie: Architecture for Capturing, Socially Transforming and
  Analyzing the Twittersphere.
\newblock In {\em Proc. of the 17th ACM Int. Conf. on Supporting Group Work},
  pages 229--238, 2012.

\bibitem{twitterecho2012}
M.~Bo\v{s}njak, E.~Oliveira, J.~Martins, E.~Mendes~Rodrigues, and L.~Sarmento.
\newblock TwitterEcho: A Distributed Focused Crawler to Support Open Research
  with Twitter Data.
\newblock In {\em Proc. of the 21st Int. Conf. on World Wide Web}, pages
  1233--1240, 2012.

\bibitem{hoarder2017}
M.~Congosto, P.~Basanta-Val, and L.~Sanchez-Fernandez.
\newblock T-Hoarder: A framework to process Twitter data streams.
\newblock {\em Journal of Network and Computer Applications}, 83:28 -- 39,
  2017.

\bibitem{dang2016}
M.~Dang and S.~Singh.
\newblock Context based interesting tweet recommendation framework.
\newblock In {\em IEEE Annual India Conference (INDICON)}, pages 1--5, 2016.

\bibitem{Dean2004}
J.~Dean and S.~Ghemawat.
\newblock {MapReduce}: Simplified Data Processing on Large Clusters.
\newblock In {\em Symposium on Operating System Design and Implementation},
  pages 10--10, 2004.

\bibitem{docker}
{Docker}.
\newblock \url{http://www.docker.com}.
\newblock [Online; accessed July, 2017].

\bibitem{GamalloSEMEVAL2014}
P.~Gamallo and M.~Garcia.
\newblock Citius: A Naive-Bayes Strategy for Sentiment Analysis on English
  Tweets.
\newblock In {\em 8th International Workshop on Semantic Evaluation (SemEval
  2014)}, pages 171--175, Dublin, Irland, 2014.

\bibitem{GamalloTASS2013}
P.~Gamallo, M.~Garcia, and S.~Fern\'andez-Lanza.
\newblock A Naive-Bayes strategy for sentiment analysis on Spanish tweets.
\newblock In {\em Workshop on Sentiment Analysis at SEPLN (TASS)}, pages
  126--132, 2013.

\bibitem{giachanou2016}
A.~Giachanou and F.~Crestani.
\newblock Like It or Not: A Survey of Twitter Sentiment Analysis Methods.
\newblock {\em ACM Comput. Surv.}, 49(2):28:1--28:41, June 2016.

\bibitem{Go2009}
A.~Go, R.~Bhayani, and L.~Huang.
\newblock Twitter sentiment classification using distant supervision.
\newblock In {\em CS224N Technical report}. University of Standford, 2009.

\bibitem{karanasou2016}
M.~Karanasou, A.~Ampla, C.~Doulkeridis, and M.~Halkidi.
\newblock Scalable and Real-Time Sentiment Analysis of Twitter Data.
\newblock In {\em Proc. of the 16th Int. Conf. on Data Mining Workshops
  (ICDMW)}, pages 944--951, 2016.

\bibitem{khuc2012}
V.~N. Khuc, C.~Shivade, R.~Ramnath, and J.~Ramanathan.
\newblock Towards Building Large-scale Distributed Systems for Twitter
  Sentiment Analysis.
\newblock In {\em Proc. of the 27th ACM Symposium on Applied Computing}, pages
  459--464, 2012.

\bibitem{Kramer2014}
J.~Kramer and C.~Gordon.
\newblock {Improvement of a Naive Bayes Sentiment Classifier Using MRS-Based
  Features}.
\newblock In {\em Joint Conf. on Lexical and Computational Semantics}, pages
  22--29, 2014.

\bibitem{tap2014}
J.~Kranjc.
\newblock {TwitterTap}.
\newblock \url{https://github.com/janezkranjc/twitter-tap}, 2014.
\newblock [Online; accessed July, 2017].

\bibitem{kumar2016}
S.~Kumar, P.~Singh, and S.~Rani.
\newblock Sentimental analysis of social media using R language and Hadoop:
  Rhadoop.
\newblock In {\em 5th Int. Conf. on Reliability, Infocom Technologies and
  Optimization (Trends and Future Directions) (ICRITO)}, pages 207--213, 2016.

\bibitem{bliu2013}
B.~Liu, E.~Blasch, Y.~Chen, D.~Shen, and G.~Chen.
\newblock Scalable sentiment classification for Big Data analysis using Naïve
  Bayes Classifier.
\newblock In {\em Proc. of the IEEE Int. Conf. on Big Data}, pages 99--104,
  2013.

\bibitem{Manning2008}
C.~Manning, P.~Raghadvan, and H.~Sch{\"u}tze.
\newblock {\em Introduction to Information Retrieval}.
\newblock Cambridge University Press, Cambridge, MA, USA, 2008.

\bibitem{MariaDB}
{MariaDB}.
\newblock \url{https://mariadb.org/}.
\newblock [Online; accessed July, 2017].

\bibitem{minanovic2014}
A.~Minanovic, H.~Gabelica, and Z.~Krstić.
\newblock Big data and sentiment analysis using KNIME: Online reviews vs.
  social media.
\newblock In {\em Proc. of the Int. Convention on Information and Communication
  Technology, Electronics and Microelectronics (MIPRO)}, pages 1464--1468,
  2014.

\bibitem{Pang2002}
B.~Pang, L.~Lee, and S.~Vaithyanathan.
\newblock {Thumbs Up?: Sentiment Classification Using Machine Learning
  Techniques}.
\newblock In {\em Conf. on Empirical Methods in Natural Language Processing},
  volume~10, pages 79--86, 2002.

\bibitem{trendminer2012}
D.~Preotiuc-Pietro, S.~Samangooei, T.~Cohn, N.~Gibbins, and M.~Niranjan.
\newblock Trendminer: An Architecture for Real Time Analysis of Social Media
  Text, 2012.

\bibitem{rahnama2014}
A.~H.~A. Rahnama.
\newblock Distributed real-time sentiment analysis for big data social streams.
\newblock In {\em Proc. of the Int. Conf. on Control, Decision and Information
  Technologies (CoDIT)}, pages 789--794, 2014.

\bibitem{Rosenthal2014}
S.~Rosenthal, P.~Nakov, A.~Ritter, and V.~Stoyanov.
\newblock {SemEval-2014 Task 9: Sentiment Analysis in Twitter}.
\newblock In {\em Int. Workshop on Semantic Evaluation}, 2014.

\bibitem{Severyn2015}
A.~Severyn, A.~Moschitti, O.~Uryupina, B.~Plank, and K.~Filippova.
\newblock Multi-lingual opinion mining on YouTube.
\newblock {\em Inform. Processing \& Management}, 52(1):46--60, 2015.

\bibitem{skuza2015}
M.~Skuza and A.~Romanowski.
\newblock Sentiment analysis of Twitter data within big data distributed
  environment for stock prediction.
\newblock In {\em Federated Conference on Computer Science and Information
  Systems (FedCSIS)}, pages 1349--1354, 2015.

\bibitem{Vilares&Alonso2015}
D.~Vilares, M.~A. Alonso, and C.~G\'omez-Rodr\'iguez.
\newblock On the usefulness of lexical and syntactic processing in polarity
  classification of Twitter messages.
\newblock {\em Journal of the American Society for Information Science},
  66(9):1799--1816, 2015.

\bibitem{Villena2013}
J.~Villena-Rom\'an, S.~Lana, E.~Mart\'inez-C\'amara, and J.~C.
  Gonz\'alez-Crist\'obal.
\newblock {TASS - Workshop on Sentiment Analysis at SEPLN}.
\newblock {\em {Procesamiento del Lenguaje Natural}}, 50:37--44, 2013.

\bibitem{wang2012}
H.~Wang, D.~Can, A.~Kazemzadeh, F.~Bar, and S.~Narayanan.
\newblock A System for Real-time Twitter Sentiment Analysis of 2012 U.S.
  Presidential Election Cycle.
\newblock In {\em Proc. of the ACL System Demonstrations}, pages 115--120,
  2012.

\bibitem{zarrad2014}
A.~Zarrad, A.~Jaloud, and I.~Alsmadi.
\newblock The Evaluation of the Public Opinion - A Case Study: MERS-CoV
  Infection Virus in KSA.
\newblock In {\em Proc. of the IEEE/ACM 7th Int. Conf. on Utility and Cloud
  Computing}, pages 664--670, 2014.

\end{thebibliography}

\end{document}